# Stabilization of two-dimensional solitons in cubic-saturable nonlinear lattices

Olga V. Borovkova, Yaroslav V. Kartashov, and Lluis Torner

*ICFO-Institut de Ciencies Fotoniques, and Universitat Politecnica de Catalunya, Mediterranean Technology Park, 08860 Castelldefels (Barcelona), Spain*

We consider soliton dynamics and stability in a nonlinear lattice formed by alternating domains with focusing cubic and saturable nonlinearities. We find that in such lattices solitons centered on cubic domains may be stabilized even in two-dimensional geometries, in spite of their intrinsic catastrophic instability in the absence of the lattice. Solitons centered on saturable domains are always unstable.

## I. Introduction

Stabilization of multidimensional solitons is a subject of continuously renewed interest. It is well-known that, in particular, two-dimensional (2D) bright solitons propagating in uniform cubic (Kerr) nonlinear media are unstable and experience catastrophic collapse [1]. Different strategies have been elucidated to achieve stabilization of such solitons, including exploitation of various types of transverse modulations of parameters of the medium, or so-called optical lattices. Thus, a shallow periodic transverse modulation of the refractive index in a material with uniform nonlinearity that creates a periodic lattice is known to play a strong stabilizing role for solitons (for recent reviews see, e.g., [2,3]) and may support not only fundamental 2D solitons, but also more complex states like bright vortex solitons or multipole solitons [4-12]. Another promising mechanism relies on the creation of purely nonlinear lattices where the nonlinearity coefficient is transversely modulated, while refractive index remains uniform [13-15]. While smooth variation of nonlinearity in the transverse plane may stabilize 2D solitons only in a very limited parameter range [13], systems with sharp step-like variations of the nonlinearity do support stable 2D solitons as it was predicted in [14], where stable radially symmetric solitons were obtained inside a nonlinear circle or annular ring embedded into a linear medium. Very recently, stability of 2D solitons was illustrated in a truly periodic nonlinear lattice in the form of array of nonlinear circles [15]. Notice that the properties of one-dimensional solitons in nonlinear lattices are particularly intriguing [16-20] and they may depart considerably from the properties of solitons in conventional lattices where the refractive index is modulated, while the nonlinearity remains uniform. Nonlinear lattices may act in combination with usual refractive index modulations [21-24], a competition that results in new soliton properties, soliton shapes and soliton mobility.

To date, stabilization of 2D solitons in the materials with inhomogeneous nonlinearity landscapes has been studied only in settings where domains with cubic nonlinearity are embedded into a linear medium. In this work we show that 2D solitons can also be stabilized in new settings, where cubic domains are embedded into materials with other types of nonlinearity, for example saturable nonlinearities. We found that in such lattices solitons centered on cubic domains may be stable in both 1D and 2D settings, while their counterparts centered on saturable domains are always unstable. Such position-dependent stability properties are in contrast to stability properties of solitons in corresponding uniform media, especially in 2D geometries, where solitons suffer from collapse in cubic materials. Solitons also exhibit power-controlled shape transformations.

## II. Theoretical model

We consider the propagation of light beams along the $\xi$-axis of the medium composed from alternating (in the transverse direction) domains with cubic and saturable nonlinearities that can be described by the nonlinear Schrödinger equation for the dimensionless field amplitude $q$:

$$i\frac{\partial q}{\partial \xi} = -\frac{1}{2}\Delta q - \frac{q|q|^2}{1+S|q|^2}, \qquad (1)$$

where $\xi$ is the propagation distance normalized to the diffraction length; the Laplace operator can be written as $\Delta = \partial^2/\partial\eta^2 + \partial^2/\partial\zeta^2$ in the case of two transverse dimensions and takes on the form $\Delta = \partial^2/\partial\eta^2$ in the case of single transverse dimension; $\eta$ and $\zeta$ are the transverse coordinates normalized to the characteristic transverse scale. We assume that the saturation parameter $S=0$ in cubic domains and $S=1$ in saturable domains. We consider two different configurations. In the setting of first type in 2D the domains with saturate nonlinearity having square cross-sections with width $w_s$ are arranged into a square array with period $w_c + w_s$, so that square saturable domains are separated by backbone-shaped cubic regions of width $w_c$ [Fig. 1(a)]. In the second setting, cubic and saturable domains were exchanged, i.e. in this case square cubic domains with width $w_c$ are arranged into square array and separated by backbone-shaped saturable regions [Fig. 1(b)]. In both cases we assume that soliton resides in the point $\eta, \zeta = 0$. The cross-sections of such structures along either $\eta$ or $\zeta$ axes represent distribution of saturable and cubic domains in corresponding 1D lattices.

## III. One-dimensional solitons

First, we consider solitons in 1D nonlinear lattices. We search for soliton solutions of Eq. (1) in the form $q = w(\eta)\exp(ib\xi)$, where the function $w(\eta)$ describes soliton profile and $b$ is the propagation constant. Representative profiles of solitons centered on the domain with cubic nonlinearity surrounded by domains with saturable nonlinearity are shown in Fig. 2(a). Such solitons always feature bell-like shapes with very weak oscillations on top of them due to transverse modulation of nonlinearity. Increasing propagation constant results in progressive localization of light inside the cubic domain. This is accompanied by monotonic growth of soliton's energy flow $U = \int_{-\infty}^{\infty} |q|^2 d\eta$ as shown by curve 1 in Fig. 2(c). Notice that energy flow vanishes when $b \to 0$, since the role of nonlinear lattice diminishes with decrease of peak amplitude. This is in contrast to soliton behavior in usual lattices, when only the refractive index is modulated, that impose restrictions on available $b$ values.

The most interesting results occur for solitons centered on saturable domains. With increase of $b$ such solitons first develop flat-top field distributions inside saturable domain, while soliton's tails inside neighboring cubic domains rapidly fade away. For even larger propagation constant values such solitons experience considerable reshaping because two humps tend to develop on soliton profile in the regions with cubic nonlinearity, so that at sufficiently high peak intensities the light concentrates almost entirely in two cubic domains nearest to the central saturable domain [see Fig. 2(b)]. This situation takes place even for small widths of domains $w_s, w_c$, although with decrease of $w_s, w_c$ such shape transformations appear at progressively increasing power levels. The tendency for localization on cubic domains can be understood if one takes into account that the nonlinear contribution to refractive index in cubic domains $\sim |q|^2$ is higher than nonlinear contribution in saturable domains $\sim |q|^2/(1+S|q|^2)$. The energy flow of solitons centered on saturable domains also monotonically increases with $b$ [see curve 2 in Fig. 2(c)]. Interestingly, while $U(b)$ curves for solitons centered on different types of domains almost coincide for $b \to 0$, for large $b$ values the energy flow of soliton centered on saturable domain is approximately two times higher than that of soliton centered on cubic domain.

Upon linear stability analysis we searched for the profiles of perturbed soliton solutions in the form $q = [w + u\exp(\delta\xi) + iv\exp(\delta\xi)]\exp(ib\xi)$, where $u, v \ll w$ are real and imaginary parts of small perturbation, while $\delta = \delta_\mathrm{r} + i\delta_\mathrm{i}$ is the perturbation growth rate that may be complex. The substitution of this expression into Eq. (1) and linearization leads to the following eigenvalue problem:

$$\begin{aligned}
\delta u = \mathcal{L}_v v = -\frac{1}{2}\Delta v + bv - v\frac{w^2 + Sw^4}{(1+Sw^2)^2}, \\
\delta v = \mathcal{L}_u u = +\frac{1}{2}\Delta u - bu + u\frac{3w^2 + Sw^4}{(1+Sw^2)^2}.
\end{aligned} \qquad (2)$$

The analysis of Eqs. (2) for solitons centered on cubic and saturable domains reveals that former type of solitons is always stable and may propagate undistorted over considerable distances even in the presence of strong perturbations, while the latter type of solitons is unstable for all propagation constant values. In Fig. 2(d) we show the dependence of real part of perturbation growth rate on propagation constant for solitons centered on saturable domains. The dependence $\delta_\mathrm{r}(b)$ is nonmonotonic and growth rate asymptotically approaches zero when $b \to \infty$. In the presence of input perturbations such solitons quickly reshape into single-peak states centered on cubic domains.

## IV. Two-dimensional solitons

Second, we address 2D solitons in settings with alternating cubic and saturable domains. Figure 3(a) shows typical profiles of solitons centered on cubic domains, while Fig. 3(b) shows solitons centered on saturable domains. One can see that upon increase of the propagation constant, solitons residing on cubic domains gradually localize in them retaining a bell-shaped distribution, while their counterparts on saturable domains exhibit considerable power-dependent shape transformations. Like in 1D, solitons in saturable domains first develop flat-top shapes, while further increase of $b$ results in appearance of four local maxima on otherwise smooth field distribution, whose positions coincide with positions of domains with cubic nonlinearity. For large $b$ values light field concentrates almost entirely in cubic regions, so that soliton profile resembles four well-localized bright spots. Notice that at low powers, when $b \to 0$ 2D solitons residing on both cubic and saturable domains extend dramatically across nonlinear lattice, without developing any noticeable shape oscillations.

In contrast to 1D geometries, the energy flow of 2D solitons in nonlinear lattice is a nonmonotonic function of $b$. In Fig. 4(a) we show the dependencies $U(b)$ for solitons centered on saturable domains for different $w_s$. In this case energy flow first grows with $b$, reaches its maximal value, and then decreases. When $b \to \infty$ the energy flow asymptotically approaches $4U_T$, where $U_T \approx 5.85$ is the energy flow of Townes soliton in uniform cubic medium. However, despite the fact that the dependence $U(b)$ is nonmonotonic and that there exist inflection point where derivative $dU/db$ changes its sign, the linear stability analysis shows that solitons centered on saturable domains in 2D case are always unstable. This is an example of system where Vakhitov-Kolokolov stability criterion may not give correct prediction of stability domains for fundamental solitons (for other examples see [13,22]). The typical evolution of unstable perturbed soliton centered on saturable domain is shown in Fig. 5(a). After rather short propagation distance the light tends to concentrate in one of cubic domains. Linear stability analysis performed with the aid of Eq. (2) allowed us to obtain the dependence of growth rate on propagation constant shown in Fig. 4(b). One can see that $\delta_r$ is nonzero in the entire soliton's existence domain, while corresponding instability is of oscillatory character. Such behavior is in contrast to well-established fact of stability of 2D solitons in uniform saturable medium, where collapse is arrested.

On the other hand, it is well-known that solitons in uniform cubic medium are unstable and may experience collapse. Still, our analysis shows that alternation of cubic and saturable nonlinear domains in the transverse plane does allow stabilization of 2D solitons centered on cubic domains in such purely nonlinear lattice. The typical dependence of energy flow $U$ on $b$ for such solitons is shown in Fig. 4(c) for two different widths of cubic domains $w_c$. Notice a completely different character of $U(b)$ dependencies for solitons centered on cubic and saturable domains. With increase of $b$ in Fig. 4(c) the energy flow decreases, reaches its minimal value at $b = b_{cr}$, and then starts increasing. Thus, there exist certain minimal value of energy flow that decreases with decrease of the width of cubic domains. Energy flow $U$ asymptotically approaches the value $U_T \approx 5.85$ when $b \to \infty$. By performing linear stability analysis we found that in the case of solitons centered on cubic domains the Vakhitov-Kolokolov stability criterion does give the correct prediction for domains of soliton stability. Namely, solitons corresponding to the branch with $dU/db > 0$ are linearly stable, while solitons from the branch $dU/db \leq 0$ are unstable. Figure 4(d) shows that the real part of the growth rate becomes zero exactly in the inflection point of $U(b)$ dependence, where derivative $dU/db$ changes its sign. The critical value of propagation constant, $b_{\mathrm{cr}}$, at which solitons become stable increases with decrease of the width of cubic domains. Direct propagation of perturbed solitons centered on cubic domains confirm the predictions of linear stability analysis. The perturbed solitons with $b > b_{\mathrm{cr}}$ clean up the noise and propagate undistorted over indefinitely long distances [see an example of stable propagation in Fig. 5(b)].

The difference in stability properties of solitons centered on cubic and saturable domains is consistent with the limited applicability of the so-called Vakhitov-Kolokolov criterion when applied to the case at hand. Thus, in addition to slope condition $dU/db > 0$ the so-called spectral stability condition should be satisfied. Such spectral instability imposes restrictions, that depend on the domain where soliton center is located, on the number of eigenvalues of linearized operators $\mathcal{L}_u, \mathcal{L}_v$ in the eigenvalue problem (2) $\delta u = \mathcal{L}_v v$, $\delta v = \mathcal{L}_u u$. While violation of the slope condition in 2D results in collapse or decay (as it occurs, for example, in uniform cubic medium), the violation of spectral stability condition results in appearance of eigenvalues with $\delta_{\mathrm{r}} \neq 0$ with associated non-symmetric eigenvectors that indicates that solitons tend to experience transverse drift under the action of such perturbations [13,17]. Thus, while in our setting the solitons residing on saturable domains satisfy slope condition for certain $b$ values, they do not satisfy spectral stability condition. As a result they tend to shift to domains with cubic nonlinearity. On intuitive physical grounds, this is understood by taking into account that in cubic domains the nonlinear contribution to refractive index is higher. At the same time the solitons residing on the cubic domains satisfy both conditions as long as $b > b_{\mathrm{cr}}$ and can be stable.

## V. Conclusions

Summarizing, the salient result of this paper is that solitons centered on cubic (Kerr) domains can be stabilized, even in 2D geometries where otherwise solitons are highly unstable in uniform media. Stabilization occurs when the soliton propagation constant exceeds a critical value. Interestingly, we also found that in both one- and two-dimensional geometries solitons centered on saturable domains of nonlinear lattice are unstable. They undergo shape transformations with increase of peak intensity when light concentrates in regions with cu-bic nonlinearity. Such position-dependent stability properties are in contrast to the soliton properties propagating in uniform media, especially in 2D geometries, a result that emphasizes the possibilities that are open by the concept of nonlinear lattices when applied to soliton science.

# Figure captions

Figure 1 (color online). Distribution of saturation parameter in two-dimensional nonlinear lattice for the case when soliton with a center in the point $\eta, \zeta = 0$ resides on saturable domain (a) and on cubic domain (b). All quantities are plotted in arbitrary dimensionless units.

Figure 2 (color online). Profiles of solitons centered on (a) cubic and (b) saturable domains. Gray regions indicate saturable domains, while white regions indicate cubic domains. (c) $U$ versus $b$ for solitons residing on cubic domain (curve 1) and on saturable domain (curve 2). (d) $\delta_r$ versus $b$ for solitons centered on saturable domain. In all cases $w_c = 1$, $w_s = 1$. All quantities are plotted in arbitrary dimensionless units.

Figure 3 (color online). Profiles of solitons centered on (a) cubic domain when $w_c = 1.5$, $w_s = 1.0$ and (b) saturable domain when $w_c = 1.0$, $w_s = 2.0$. All quantities are plotted in arbitrary dimensionless units.

Figure 4 (color online). (a) $U$ versus $b$ for solitons centered on domain with saturable nonlinearity at $w_s = 1.5$ (curve 1), 2.0 (curve 2), and 2.5 (curve 3). (b) $\delta_r$ versus $b$ for solitons centered on domain with saturable nonlinearity at $w_s = 1.5$ (curve 1) and 2.5 (curve 2). In (a) and (b) $w_c = 1.0$. (c) $U$ versus $b$ for solitons centered on domain with cubic nonlinearity at $w_c = 1.0$ (curve 1) and 1.5 (curve 2). Circles in (c) separate stable and unstable branches. (d) $\delta_r$ versus $b$ for solitons centered on domain with cubic nonlinearity at $w_c = 1.5$. In (c) and (d) $w_s = 1.0$. All quantities are plotted in arbitrary dimensionless units.

Figure 5 (color online). (a) Decay of soliton centered on saturable domain at $b = 2.0$, $w_c = 1.0$, $w_s = 2.0$, and (b) stable propagation of soliton centered on cubic domain at $b = 2.0$, $w_c = 1.5$, $w_s = 1.0$. Field modulus distributions are shown at different distances. All quantities are plotted in arbitrary dimensionless units.

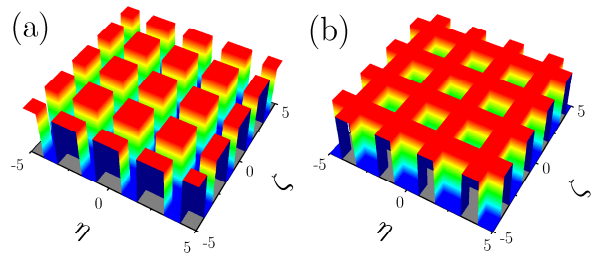

Figure 1 (color online). Distribution of saturation parameter in two-dimensional nonlinear lattice for the case when soliton with a center in the point $\eta, \zeta = 0$ resides on saturable domain (a) and on cubic domain (b).

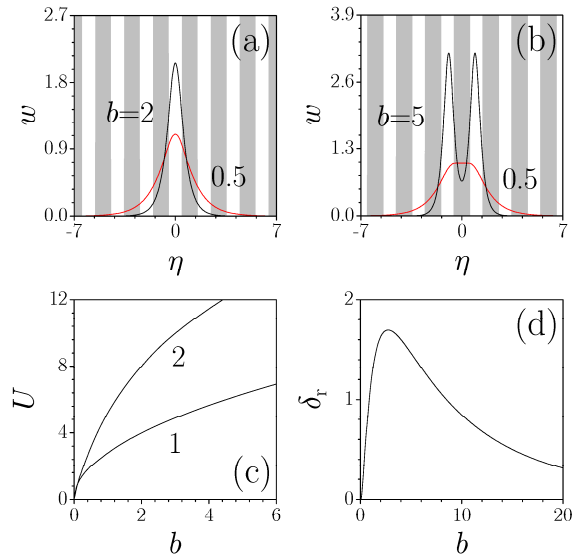

Figure 2 (color online). Profiles of solitons centered on (a) cubic and (b) saturable domains. Gray regions indicate saturable domains, while white regions indicate cubic domains. (c) $U$ versus $b$ for solitons residing on cubic domain (curve 1) and on saturable domain (curve 2). (d) $\delta_{\rm r}$ versus $b$ for solitons centered on saturable domain. In all cases $w_{\rm c}=1$, $w_{\rm s}=1$. All quantities are plotted in arbitrary dimensionless units.

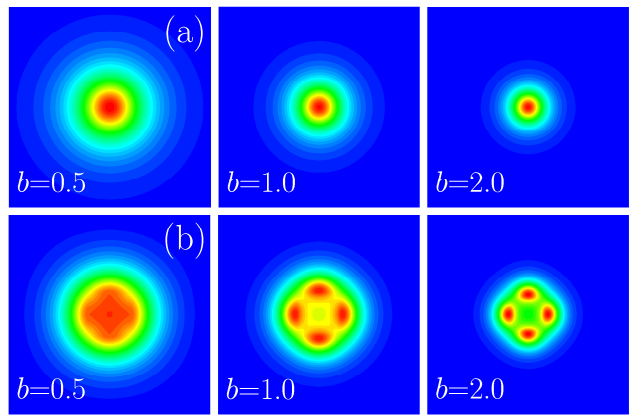

Figure 3 (color online). Profiles of solitons centered on (a) cubic domain when $w_c = 1.5$, $w_s = 1.0$ and (b) saturable domain when $w_c = 1.0$, $w_s = 2.0$. All quantities are plotted in arbitrary dimensionless units.

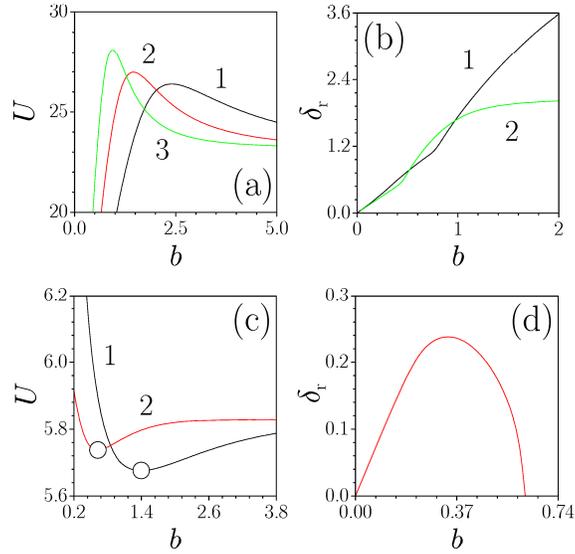

Figure 4 (color online). (a) $U$ versus $b$ for solitons centered on domain with saturable nonlinearity at $w_s = 1.5$ (curve 1), 2.0 (curve 2), and 2.5 (curve 3). (b) $\delta_r$ versus $b$ for solitons centered on domain with saturable nonlinearity at $w_s = 1.5$ (curve 1) and 2.5 (curve 2). In (a) and (b) $w_c = 1.0$. (c) $U$ versus $b$ for solitons centered on domain with cubic nonlinearity at $w_c = 1.0$ (curve 1) and 1.5 (curve 2). Circles in (c) separate stable and unstable branches. (d) $\delta_r$ versus $b$ for solitons centered on domain with cubic nonlinearity at $w_c = 1.5$. In (c) and (d) $w_s = 1.0$. All quantities are plotted in arbitrary dimensionless units.

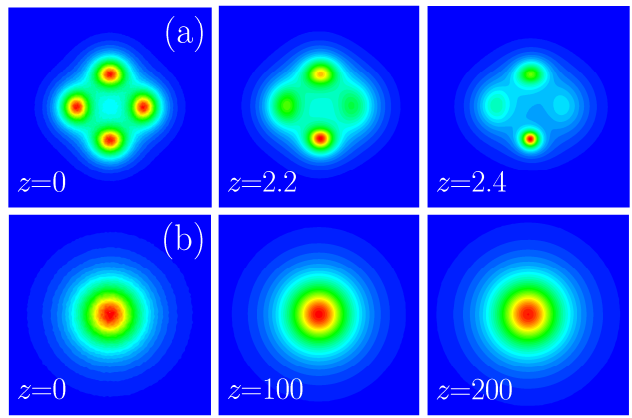

Figure 5 (color online). (a) Decay of soliton centered on saturable domain at $b = 2.0$, $w_\text{c} = 1.0$, $w_\text{s} = 2.0$, and (b) stable propagation of soliton centered on cubic domain at $b = 2.0$, $w_\text{c} = 1.5$, $w_\text{s} = 1.0$. Field modulus distributions are shown at different distances. All quantities are plotted in arbitrary dimensionless units.